\documentclass[a4paper, 12pt]{article}

\usepackage[utf8]{inputenc}
\usepackage[T1]{fontenc}
\usepackage[english]{babel}
\usepackage{lmodern} 
\usepackage{graphicx}
\usepackage{microtype}

\usepackage{bbm}
\usepackage{amsmath}
\usepackage{amsthm}
\usepackage{amssymb}
\usepackage{mathrsfs}
\usepackage[babel=true]{csquotes}
\usepackage{verbatim}
\usepackage{ytableau}
\usepackage{cite}

\usepackage{hyperref}

\numberwithin{equation}{section}

\begin{document}

\begin{titlepage}

\setcounter{page}{1}

\begin{center}

\hfill


{\LARGE Induced Action for Conformal Higher Spins from Worldline Path Integrals}


\vskip 30pt

{\sc Roberto Bonezzi\,$^a$}

\vskip 30pt

{\hskip -.1truecm Groupe de M\'ecanique et Gravitation,
Unit of Theoretical and Mathematical Physics,\\
University of Mons-- UMONS, 20 place du Parc, 7000 Mons, Belgium}

\vskip 10pt

\end{center}

\vskip 30pt

\paragraph{Abstract.} 
Conformal higher spin (CHS) fields, despite being non unitary, provide a remarkable example of a consistent interacting higher spin theory in flat space background, that is local to all orders. The non-linear action is defined as the logarithmically UV divergent part of a one-loop scalar effective action. In this paper we take a particle model, that describes the interaction of a scalar particle to the CHS background, and compute its path integral on the circle. We thus provide a worldline representation for the CHS action, and rederive its quadratic part. We plan to come back to the subject, to compute cubic and higher vertices, in a future work.

\vspace{1.5cm}
\begin{flushleft} \footnotesize
{${}^a$ \href{mailto:roberto.bonezzi@umons.ac.be}{roberto.bonezzi@umons.ac.be}}

\end{flushleft}

\end{titlepage}


\newpage

{
\tableofcontents }       

\section{Introduction}

Four dimensional Maxwell ${(s=1)}$ theory is the first known example of a conformally invariant physical system. Similarly, massless matter lagrangians (${s=0,\,1/2}$) have conformal symmetry in flat space, that can be enhanced to general covariance plus local Weyl symmetry when coupled to a curved spacetime metric. On the other hand, for spin greater than one ordinary two-derivative theories, such as (super)gravities (${s=2,\,3/2}$) and massless higher spin theories (${s>2}$), are not conformal. Weyl squared gravity, with higher derivative lagrangian ${{\cal L}=\sqrt{g}(W_{\mu\nu\rho\sigma})^2\approx h^{\mu\nu}\Box^2h_{\mu\nu}+\ldots\,}$, and its supersymmetric extensions \cite{Kaku:1978nz,Bergshoeff:1980is,Fradkin:1985am} are alternative models for ${s\leq2}$ possessing local Weyl symmetry besides diffeomorphism invariance, and hence rigid conformal symmetry around flat space.
Conformal higher spin fields (CHS) \cite{Fradkin:1985am,Siegel:1988gd,Fradkin:1989md,Tseytlin:2002gz,Segal:2002gd,Shaynkman:2004vu,Marnelius:2008er,Vasiliev:2009ck,Bekaert:2010ky,Bandos:2011wi,Bekaert:2012vt,Haehnel:2016mlb} are the ${s>2}$ generalization of the Weyl graviton and conformal gravitino\footnote{For superconformal HS theories, see also \cite{Kuzenko:2016qdw,Kuzenko:2017ujh}. }. In four dimensional flat space they are described by the free lagrangian\footnote{We will discuss only bosonic totally symmetric fields. In arbitrary even dimensions one has to add a power $\Box^\frac{d-4}{2}$ of the laplacian.}
\begin{equation}\label{FreeCHS}
S[h]=\sum_s\int d^4x\,h_s\,P_s\,\Box^s\,h_s\;,
\end{equation}
where ${h_s=h_{\mu_1...\mu_s}}$ and $P_s$ is the spin-$s$ transverse-traceless projector built out of $s$ powers of ${P_1:=\delta^\mu_\nu-\frac{\partial^\mu\partial_\nu}{\Box}}\,$.
The above action thus describes pure spin $s$ states (transverse and traceless) off-shell, and is invariant under differential and algebraic gauge transformations:
\begin{equation}\label{GT}
\delta h_s=\partial\epsilon_{s-1}+\eta\,\alpha_{s-2}
\end{equation}
generalizing linearized diffeomorphisms and Weyl symmetry of conformal (super)gravity. The higher derivative\footnote{One can describe CHS dynamics with ordinary two-derivative lagrangians, at the expense of introducing auxiliary fields \cite{Metsaev:2007fq,Metsaev:2007rw}.} kinetic operator ensures locality of the action, at the price of formally loosing unitarity. Contrary to the case of free massless higher spins, that propagate only on maximally symmetric backgrounds, there is evidence \cite{Nutma:2014pua,Grigoriev:2016bzl,Beccaria:2017nco} that conformal higher spins can propagate consistently on Bach-flat backgrounds, \emph{i.e.} on the equations of motion of Weyl squared gravity. In fact, quite remarkably, the above action and linear gauge symmetry admit a consistent fully non-linear completion \cite{Tseytlin:2002gz,Segal:2002gd,Bekaert:2010ky}, that is well defined around flat space and local to all orders in the fields. Indeed, unlike the case of massless higher spins, the absence of dimensionful parameters fixes the number of derivatives of each vertex uniquely.\footnote{In arbitrary even dimension $d$ the conformal weight of $h_s$ is $2-s\,$. Given an $n$th order vertex with fields of spin $(s_1,...,\,s_n)\,$, the number of derivatives is fixed to $N=d+\sum_{i=1}^ns_i-2n\,$.} The CHS theory is power-counting renormalizable but, since Weyl and higher order algebraic symmetries are gauged, it has to be free of conformal and higher spin anomalies in order to be consistent at the quantum level. In the low spin case $s\leq2\,$, vanishing of the total Weyl anomaly can be achieved only by ${{\cal N}=4}$ conformal supergravity coupled to four ${{\cal N}=4}$ SYM multiplets \cite{Fradkin:1981jc,Fradkin:1983tg}. In the case of four dimensional CHS with one field of each integer spin, the a-coefficient\footnote{In four dimensions the Weyl anomaly contains only two relevant structures: the Euler density, whose coefficient is usually named $a$, and the square of the Weyl tensor, whose coefficient is $c\,$.} of the Weyl anomaly vanishes upon a (regularized) summation over all spins \cite{Giombi:2013yva,Tseytlin:2013jya}. Similarly, by a less straightforward argument, the c-coefficient seems to vanish as well \cite{Tseytlin:2013jya,Giombi:2014iua,Beccaria:2014xda,Beccaria:2015vaa,Beccaria:2017nco,Beccaria:2017lcz}, while hints for the absence of anomalies in the higher spin algebraic symmetries rely only on symmetry considerations.

Besides being interesting on its own, as it gives a nontrivial example of an interacting higher spin theory in flat space, CHS fields are intimately related to massless higher spin theories in Anti de Sitter space \cite{Vasiliev:1990en,Vasiliev:1990vu,Vasiliev:1992av,Vasiliev:1999ba,Vasiliev:2003ev,Bekaert:2005vh,Didenko:2014dwa} via the vectorial AdS/CFT correspondence \cite{Sezgin:2002rt,Klebanov:2002ja,Giombi:2009wh,Giombi:2010vg,Giombi:2012ms,Giombi:2016ejx}. Moreover, the non-linear CHS action naturally arises as an induced action \cite{Liu:1998bu,Tseytlin:2002gz,Segal:2002gd,Bekaert:2010ky} in the holographic context: The free CFT of $N$ complex scalars admits an infinite number of conserved conformal currents of every spin in the $U(N)$ singlet sector, $J_s\sim\phi^*_i\partial^s\phi^i\,$. The dual fields to these conformal currents are identified with massless higher spin gauge fields in AdS space, whose boundary values $h_s$ source the $J_s$ currents and can in turn be seen as CHS fields on the boundary.
The scalar path integral with sources $\sum_sJ_s\,h_s$ yields the generating functional $\Gamma[h]$ of correlators of the conformal currents and, according to AdS/CFT correspondence, should be equal to the on-shell value of the, yet unknown\footnote{Vasiliev's equations lack a standard variational principle. Non-standard actions of covariant hamiltonian type have been proposed in \cite{Boulanger:2011dd,Boulanger:2012bj,Boulanger:2015kfa,Bonezzi:2015igv,Bonezzi:2016ttk} . From an holographic perspective, CFT correlators have been used to reconstruct AdS vertices in \cite{Bekaert:2014cea,Bekaert:2015tva,Sleight:2016dba}.}, action of massless higher spins in AdS\footnote{Direct matching of free gauge theory correlators with AdS Witten diagrams has been investigated in \cite{Gopakumar:2003ns,Gopakumar:2004qb,Gopakumar:2005fx} in order to exploit open-closed string duality. In particular, in \cite{Gopakumar:2003ns} one-loop open string diagrams in the field theory limit (hence worldline loops) were shown to reproduce tree level diagrams in AdS by direct change of variables in the moduli space.}. However, the same generating functional $\Gamma[h]$ can be interpreted\footnote{In the standard AdS/CFT context \cite{Maldacena:1997re,Gubser:1998bc,Witten:1998qj} the boundary values of bulk fields are fixed, non dynamical sources for CFT correlators. From a pure boundary perspective, however, one can see the coupling $\sum_sJ_s\,h_s$ as a Noether coupling that gauges the infinite dimensional symmetry algebra \cite{Eastwood:2002su,Vasiliev:2003ev} generated by the charges associated to the currents $J_s\,$. Moreover, even in the AdS/CFT context one can give different, Neumann type, boundary conditions to bulk fields, allowing them to fluctuate on the boundary \cite{Balasubramanian:2000pq,Compere:2008us,Giombi:2013yva}.}, from a pure boundary viewpoint, as a one-loop effective action for the CHS fields $h_s\,$, that inherit the linearized gauge symmetry \eqref{GT} thanks to conservation and tracelessness of the currents $J_s\,$. The logarithmically divergent part of $\Gamma[h]$ is local and gauge invariant and can be thus identified as the classical non-linear action $S_{\rm CHS}[h]$ for conformal higher spins \cite{Segal:2002gd,Bekaert:2010ky}. 

The aim of this paper is to construct a quantum mechanical path integral to represent the effective action $\Gamma[h]\,$. Since the coupling $\sum_sJ_s\,h_s$ is quadratic in the scalar fields, $\Gamma[h]$ is given by the functional determinant
\begin{equation}\label{H}
\Gamma[h]=N\,\log\,{\rm Det}[-\Box+\hat H]\;,
\end{equation}
where $\hat H$ is a differential operator linear in the CHS fields. Such type of one-loop effective actions is the most suitable to be computed by using first-quantized worldline models \cite{Schubert:2001he,Bern:1991aq,Strassler:1992zr,Bastianelli:1992ct,DHoker:1995uyv,Reuter:1996zm,Bastianelli:2002fv,Bastianelli:2002qw,Bastianelli:2005vk,Dai:2008bh,Bastianelli:2013tsa,Bastianelli:2013pta}. For instance, free massless scalar particles are described in first quantization by the relativistic worldline action
\begin{equation}
S[x,p,e]=\int_0^1d\tau\,\Big[p_\mu\dot x^\mu-\tfrac{e}{2}\,p^2\Big]\quad\Leftrightarrow\quad S[x,e]=\int_0^1d\tau\,\frac{\dot x^2}{2e}\;,
\end{equation}
where $e(\tau)$ is the einbein, that enforces the mass-shell constraint $p^2\approx0$ and, equivalently, ensures local $\tau$-reparametrization invariance. Coupling to a background curved metric $g_{\mu\nu}(x)$ and $U(1)$ gauge field $A_\mu(x)$ can be readily achieved by
\begin{equation}\label{S12}
\begin{split}
S_{g,A}[x,p,e] &= \int_0^1d\tau\,\Big[p_\mu\dot x^\mu-\tfrac{e}{2}\,g^{\mu\nu}(p_\mu-A_\mu)(p_\nu-A_\nu)\Big]\quad\Leftrightarrow\quad\\ S_{g,A}[x,e] &= \int_0^1d\tau\,\Big[\tfrac{1}{2e}\,g_{\mu\nu}\dot x^\mu\dot x^\nu+A_\mu\dot x^\mu\Big]\;.
\end{split}
\end{equation}
Quantization of the above actions on the circle gives the scalar loop contribution to the QFT one-loop effective action for gravitons and photons. To be precise, when the hamiltonian is \emph{not} of the form ${H=p^2+V(x)}\,$, as it is the case in curved spacetime \cite{Bastianelli:2006rx}, the naive classical action does not give the correct quantum amplitudes, due to ordering issues in the quantum hamiltonian, and a local counterterm has to be added to the action \eqref{S12} before using it in the path integral.
One can add spinning degrees of freedom to the quantum particle \cite{Gershun:1979fb,Henneaux:1987cp,Howe:1988ft,Kuzenko:1995mg,Bastianelli:2008nm,Corradini:2010ia,Bastianelli:2009eh,Bastianelli:2012bn}, in order to give contributions of fields with nonzero spin in the loop. In the present case, since we are interested in the effective action $\Gamma[h]$ generated by a scalar loop, the scalar particle example will suffice. 

In order to describe the interaction of the relativistic particle to background CHS fields, we shall employ the action proposed in \cite{Segal:2001di}, \emph{i.e.}
\begin{equation}\label{Segal}
S_h[x,p,e]=\int_0^1d\tau\,\Big[p_\mu\dot x^\mu-e\,G(x,p)\Big]\;,
\end{equation}
where in the generalized hamiltonian ${G(x,p)=p^2+{\cal H}(x,p)}$ the conformal higher spin fields, contained in the $p$-power series expansion of ${\cal H}(x,p)\,$, are treated as perturbations over the flat space background $p^2\,$. In the low spin example \eqref{S12} we gave the expression for the action both in phase space and configuration space. In most worldline applications one employs the configuration space action, but in the case at hand it seems much more convenient to stay with the phase space action and to perform the path integral directly in phase space.\footnote{For very similar reasons, phase space worldline path integrals have been used in \cite{Bonezzi:2012vr,Ahmadiniaz:2015qaa} in the context of non-commutative field theory.} Indeed, the arbitrary dependence on momenta of ${\cal H}(x,p)$ makes the inversion $p=p(\dot x)$ quite cumbersome, along with the appearence of inverse powers of $\dot x^2$ that would produce singularities in perturbation theory. The issue of quantum ordering of the operator $\hat H$ of \eqref{H} in relation to the classical interaction vertex ${\cal H}(x,p)$ will be discussed in the main text.

In the next section we start by reviewing the construction of \cite{Segal:2002gd,Bekaert:2010ky}, that allows to find the explicit form of the operator $\hat H\,$. We proceed by introducing the above worldline model and discuss its symmetries. Finally, we quantize the action \eqref{Segal} by computing explicitly the path integral on the circle. By doing so we end up with a Scwhinger proper time representation of the effective action $\Gamma[h]$ that allows to extract the logarithmic divergence defining the classical action $S_{\rm CHS}[h]\,$. For illustrative purpose we shall rederive the quadratic action \cite{Fradkin:1985am,Segal:2002gd,Bekaert:2010ky}, while we plan to address cubic and higher vertices in a future work.
We conclude in Section \ref{Discussion}
by pointing out some aspects of the present formalism that may be improved, and discussing some interesting directions for future investigations.

\section{Induced action for Conformal Higher Spins}

Let us start by considering a massless complex scalar field in flat spacetime of even dimension $d\,$, with action
\begin{equation}
S_0[\phi]=\int d^dx\,\partial^\mu\phi^*\partial_\mu\phi\;.
\end{equation}
Being a free theory, it possesses an infinite number of conserved currents $J_{\mu(s)}=\phi^*\partial_{\mu_1}...\partial_{\mu_s}\phi+...$\footnote{Indices denoted with the same letter and groups of indices $\mu(k)$ are intended as symmetrized with strength one, \emph{e.g.} $J_{\mu(s)}:=J_{(\mu_1...\mu_s)}\,$.} of arbitrary integer spin $s=0,1,2,...$ and conformal dimension $\Delta_{J_s}=d-2+s\,$, that can be made traceless thanks to conformal invariance \cite{Craigie:1983fb,Berends:1985xx}. Conservation $\partial^\nu J_{\nu\mu(s-1)}\approx0$ and tracelessness $J^\alpha{}_{\alpha\mu(s-2)}\approx0$  hold on the scalar mass-shell $\Box\phi\approx0\,$.
In this setting one can introduce conformal higher spin fields (CHS) via the Noether interactions
\begin{equation}\label{Noether}
S_{\rm int}[\phi,h]=\sum_{s=0}^\infty\frac{(i)^s}{s!}\int d^dx\,J^{\mu(s)}\,h_{\mu(s)}\;,
\end{equation}
that are invariant, on the free field equations $\Box\phi\approx0\,$, under the gauge transformations
\begin{equation}\label{gaugelin}
\delta_{\rm lin} h_{\mu(s)}=\partial_\mu\varepsilon_{\mu(s-1)}+\eta_{\mu\mu}\,\alpha_{\mu(s-2)}\;,
\end{equation}
that are the linearized higher spin generalization of the gauge symmetries of conformal gravity. These on-shell symmetries can be deformed to full off-shell ones leaving invariant the total action
\begin{equation}
S[\phi,h]=S_0[\phi]+S_{\rm int}[\phi,h]\;,
\end{equation}
by supplementing both the gauge fields and the scalar with extra transformations of the form
\begin{equation}
\delta\phi={\cal O}(\phi)\;,\quad \delta h_s=\delta_{\rm lin}h_s+{\cal O}(h)\;.
\end{equation}
The UV logarithmically divergent part\footnote{The logarithmic divergence is present only in even dimensions, that is the only case we will treat here.} of the effective action $\Gamma[h]\,$, induced by the scalar path integral
\begin{equation}\label{scalarPI}
e^{-\Gamma[h]}=\int{\cal D}\phi^*{\cal D}\phi\,e^{-S[\phi,h]}={\rm Det}^{-1}\left(-\Box+\hat {H}\right)\;,
\end{equation}
is local\footnote{The induced action contains vertices with arbitrary powers of higher spin fields but, due to the absence of dimensionful parameters, the number of derivatives is bounded by the number of fields and sum of the spins involved.} and invariant under the full transformation $\delta h_s=\partial\varepsilon_{s-1}+\eta\,\alpha_{s-2}+{\cal O}(h)\,$. It can thus be used to define a fully non-linear classical action for conformal higher spin fields, and at the quadratic level it has been shown to reproduce the free action of \cite{Fradkin:1985am}. In \eqref{scalarPI} $\hat{H}$ is a differential operator linear in $h_s\,$, whose precise form will be now reviewed following \cite{Bekaert:2010ky}.

\subsection{Noether interaction and symmetries}

The generating function of all the traceless conserved currents $J_{\mu(s)}$
\begin{equation}\label{generatingJ}
J(x,u):=\sum_{s=0}^\infty\frac{1}{s!}\,J_{\mu_1...\mu_s}(x)\,u^{\mu_1}...\,u^{\mu_s}=\sum_{s=0}^\infty J_s(x,u)
\end{equation}
obeying $\partial_u\cdot\partial_x J(x,u)\approx0$ and $\partial^2_uJ(x,u)\approx0$
can be written as \cite{Bekaert:2010ky}
\begin{equation}
J(x,u)=\Pi_d\,{\cal J}(x,u)\;,\quad {\cal J}(x,u):=\phi^*(x+u/2)\phi(x-u/2)\;,
\end{equation}
where ${\cal J}(x,u)$ generates traceful conserved currents, that are mapped to the traceless ones by the operator
\begin{equation}
\Pi_d:=\sum_{n=0}^\infty\frac{1}{n!(-\hat N-\tfrac{d-5}{2})_n}\,\left[\frac{\partial^{2}-g\,\Box}{16}\right]^n\;,
\end{equation}
with the Pochhammer symbol defined by $(a)_n:=\frac{\Gamma(a+n)}{\Gamma(a)}\,$, and  
\begin{equation}
\hat N:=u\cdot\partial_u\;,\quad\partial:=u\cdot\partial_x\;,\quad g:=u^2\;.
\end{equation}
In terms of the higher spin generating function
\begin{equation}
h(x,u):=\sum_{s=0}^\infty\frac{1}{s!}\,h_{\mu_1...\mu_s}(x)\,u^{\mu_1}...\,u^{\mu_s}=\sum_{s=0}^\infty h_s(x,u)\;,
\end{equation}
the Noether interaction \eqref{Noether} can be written as
\begin{equation}\label{NoetherRewritten}
\begin{split}
S_{\rm int}[\phi, h] &= \int d^dx\,J(x,i\partial_u)\,h(x,u)\rvert_{u=0}=\int d^dx\,{\cal J}(x,i\partial_u)\,{\cal H}(x,u)\rvert_{u=0}\\
&=\int d^dx\,e^{i\partial_u\cdot\partial_v}{\cal J}(x,v)\,{\cal H}(x,u)\rvert_{u,v=0}\;,
\end{split}
\end{equation}
where the transformed generating function of the gauge fields ${\cal H}(x,u)$ is obtained upon integrating by parts the spacetime derivatives in $\Pi_d\,$, and reads
\begin{equation}\label{gaugefieldredef}
{\cal H}(x,u)={\cal P}_d\,h(x,u)\;,\quad {\cal P}_d:=\sum_{n=0}^\infty\frac{1}{n!(\hat N+n+\tfrac{d-3}{2})_n}\left[\frac{\partial^{*2}-{\rm Tr}\,\Box}{16}\right]^n\;,
\end{equation}
the inverse map being given by \cite{Bekaert:2010ky}
\begin{equation}\label{inverseP}
h(x,u)={\cal P}^{-1}_d\,{\cal H}(x,u)\;,\quad {\cal P}^{-1}_d:=\sum_{n=0}^\infty\frac{(-1)^n}{n!(\hat N+\tfrac{d-1}{2})_n}\left[\frac{\partial^{*2}-{\rm Tr}\,\Box}{16}\right]^n\;,
\end{equation}
where we defined the divergence $\partial^*:=\partial_u\cdot\partial_x$ and trace ${\rm Tr}:=\partial_u^2$ operators. Despite the infinite series appearing in \eqref{gaugefieldredef}, each spin-$s$ component of the conformal fields $h_s$ produces a finite tail of traces and divergences, as it can be seen by rewriting
\begin{equation}\label{CHSbasis}
{\cal H}(x,u)=\sum_{s=0}^\infty\sum_{n=0}^{[s/2]}\frac{1}{n!(s-n+\tfrac{d-3}{2})_n}\left[\frac{\partial^{*2}-{\rm Tr}\,\Box}{16}\right]^n\,h_s(x,u)\;.
\end{equation}
By introducing the Fourier transform of ${\cal J}(x,v)$ in $v$-space:
\begin{equation}
{\cal J}(x,v)=\int\frac{d^dp}{(2\pi)^d}\,e^{-iv\cdot p}\rho(x,p)\;,
\end{equation}
the interaction \eqref{NoetherRewritten} can be further rewritten in the form
\begin{equation}\label{NoetherRewritten2}
S_{\rm int}[\phi,h]=\int\frac{d^dxd^dp}{(2\pi)^d}\,\rho(x,p)\,{\cal H}(x,p)\;.
\end{equation}
In \cite{Bekaert:2009ud} it has been shown that, upon introducing the first quantized Hilbert space where $x^\mu$ and $-i\frac{\partial}{\partial x^\mu}$ realize the algebra $[\hat X^\mu,\hat P_\nu]=i\,\delta^\mu_\nu$ and whereby the field $\phi(x)$ can be written as the wave function $\langle x\vert\phi\rangle\,$, the ``density matrix'' $\rho(x,p)$ is the Weyl symbol of  the operator $\vert\phi\rangle\langle\phi\vert\,$. This allows, using the standard tools of Weyl quantization \cite{Weyl:1927vd,Wigner:1932eb,Moyal:1949sk}, to finally cast the action \eqref{NoetherRewritten2} as the inner product
\begin{equation}\label{ActionBraket}
S_{\rm int}[\phi,h]={\rm Tr}\left[\vert\phi\rangle\langle\phi\vert\hat H\right]=\langle\phi\vert\hat H\vert\phi\rangle\;,
\end{equation}
where $\hat H(\hat X,\hat P)$ is the operator with Weyl symbol given by ${\cal H}(x,p)\,$, \emph{i.e.}
\begin{equation}\label{WeylOrdering}
\hat H(\hat X,\hat P)=\int\frac{d^dxd^dp}{(2\pi)^d}{\cal H}(x,p)\,\int\frac{d^dyd^dk}{(2\pi)^d}\,e^{ik\cdot(x-\hat X)-iy\cdot(p-\hat P)}\;.
\end{equation}
The total action entering the path integral \eqref{scalarPI} can thus be written as
\begin{equation}
S[\phi,h]=\langle\phi\vert\hat P^2+\hat H(\hat X,\hat P)\vert\phi\rangle\;,
\end{equation}
which is clearly invariant under
\begin{equation}\label{ExactSymm}
\vert\phi\rangle\;\rightarrow\;\hat O^{-1}\vert\phi\rangle\;,\quad (\hat P^2+\hat H)\;\rightarrow\;\hat O^\dagger(\hat P^2+\hat H)\hat O\;.
\end{equation}
In terms of the hermitian operators $\hat E$ and $\hat A$ defined by $\hat O={\rm exp}(\hat A+i\,\hat E)\,$, the infinitesimal gauge transformations of the CHS fields contained in $\hat H$ are given by
\begin{equation}
\delta\hat H=i\,[\hat P^2+\hat H,\hat E]+\{\hat P^2+\hat H,\hat A\}=i\,[\hat P^2,\hat E]+\{\hat P^2,\hat A\}+{\cal O}(h)\;,
\end{equation}
the linearized transformations \eqref{gaugelin} descending from the action of the $\hat P^2$ part. The symmetries associated to $\hat E$ correspond to the differential ones $\delta_\epsilon h_s=\partial\epsilon_{s-1}+...$ and are preserved at the quantum level, while the generator $\hat A\,$, corresponding to the generalized Weyl symmetry $\delta_\alpha h_s=\eta\,\alpha_{s-2}+...\,$, develops a quantum anomaly due to the non invariant measure of the path integral \eqref{scalarPI}. Nonetheless, it can be shown \cite{Bekaert:2010ky} that the UV logarithmically divergent part of the effective action preserves the full symmetry generated by $\hat A+i\hat E\,$, and can indeed be identified with the conformal higher spin action.

\subsection{Effective action and worldline path integral}

The effective action $\Gamma[h]$ is given by
\begin{equation}\label{GammaSchwinger}
\Gamma[h]={\rm Tr}\,{\rm log}\left(\hat P^2+\hat H\right)=-\int_0^\infty\frac{dT}{T}\,{\rm Tr}\left[e^{-T(\hat P^2+\hat H)}\right]
\end{equation}
in Schwinger proper time representation. 
The trace of the heat kernel
\begin{equation}
K[T;h]:={\rm Tr}\left[e^{-T(\hat P^2+\hat H)}\right]
\end{equation}
admits a Laurent expansion in powers of $T$ when the higher spins in $\hat H$ are treated as a perturbation over the flat spin two background $\hat P^2\,$.
Accordingly, upon introducing a cut-off $\Lambda$ in the small-$T$ region, the UV-regulated effective action can be organized according to its divergencies as
\begin{equation}
\Gamma_\Lambda[h]:=-\int_{\frac{1}{\Lambda^2}}^\infty\frac{dT}{T}\,K[T;h]=\sum_{n=1}^\infty\Lambda^{2n}\,\Gamma_n[h]+{\rm log}\Lambda\,S_{\rm CHS}[h]+\Gamma_{\rm fin}[h]+{\cal O}(\Lambda^{-2})\;,
\end{equation}
where the local, gauge invariant coefficient of the logarithmic divergence defines the conformal higher spin action being looked for.

As the heat kernel $K[T;h]$ can be viewed as the trace of the (euclidean) evolution operator associated to the quantum mechanical Hamiltonian $\hat P^2+\hat H(\hat X,\hat P)\,$, it is natural to represent it via a first quantized path integral. More so, the entire effective action \eqref{GammaSchwinger} arises from the quantization of a relativistic particle model \cite{Segal:2001di} that we briefly discuss. Consider the point particle hamiltonian action
\begin{equation}\label{Particletauinv}
S[x,p;e]=\int_0^1d\tau\left[p_\mu\dot x^\mu-e\,G(x,p)\right]\;,
\end{equation}
where
\begin{equation}
G(x,p)=p^2+{\cal H}(x,p)\approx0
\end{equation}
is the generalized mass-shell constraint imposed by the Lagrange multiplier $e(\tau)\,$, that is associated with $\tau$-reparametrization invariance under
\begin{equation}\label{Localtauinv}
\delta x^\mu=\xi\{x^\mu,G\}_{\rm P.B.}\;,\quad \delta p_\mu=\xi\{p_\mu,G\}_{\rm P.B.}\;,\quad \delta e=\dot\xi\;,
\end{equation}
where $\{,\}_{\rm P.B.}$ denotes the Poisson bracket and $\xi(\tau)$ is a worldline local parameter. The action \eqref{Particletauinv} describes the propagation of a relativistic spinless particle in the background of the CHS fields contained in ${\cal H}$ according to \eqref{CHSbasis}. The Lagrange multiplier $e(\tau)\,$, called einbein, is the gauge field for $\tau$-reparametrizations, and it can be viewed as an intrinsic frame field on the worldline. The infinitesimal gauge transformations of the background fields, generated by
\begin{equation}
\delta_\epsilon{\cal H}(x,p)=\{\epsilon(x,p),p^2+{\cal H}(x,p)\}_{\rm P.B.}\;,
\end{equation}
leave the action invariant when accompanied by the phase space transformations
\begin{equation}
\delta_\epsilon x^\mu=\{x^\mu,\epsilon(x,p)\}_{\rm P.B.}\;,\quad \delta_\epsilon p_\mu=\{p_\mu,\epsilon(x,p)\}_{\rm P.B.}\;.
\end{equation}
This is the first quantized realization of the $\hat E$-type symmetries discussed in the field theory language, while the counterpart of the generalized Weyl symmetries $\hat A$ can be viewed as the invariance of the constraint surface $G(x,p)\approx0$ under $G(x,p)\;\rightarrow\; e^{\alpha(x,p)}\,G(x,p)\,$. The action is indeed invariant under the combined transformations
\begin{equation}\label{ParticleWeyl}
\delta_\alpha{\cal H}(x,p)=\alpha(x,p)\,\Big(p^2+{\cal H}(x,p)\Big)\;,\quad\delta_\alpha e=-\alpha(x,p)\,e
\end{equation}
and, as we shall see next, the transformation of the einbein is responsible for breaking the $\alpha$-symmetry at the quantum level.

As it is well known from the  cases of interaction with scalar, vector and gravitational backgrounds, the effective action $\Gamma[h]$ can be obtained by quantizing the action \eqref{Particletauinv} on the circle:
\begin{equation}
\Gamma[h]=\int_{S^1}\frac{DxDpDe}{\rm VolGauge}\,e^{-S_E[x,p;e]}\;,
\end{equation}
where division by the gauge group volume entails the gauge fixing procedure for the local $\tau$-reparametrizations, and $S_E$ denotes the euclidean version of the action \eqref{Particletauinv}, \emph{i.e.}
\begin{equation}\label{ParticletauinvEuc}
S_E[x,p;e]=\int_0^1d\tau\left[-i\,p_\mu\dot x^\mu+e\,G(x,p)\right]\;.
\end{equation}
As mentioned in the Introduction, when the quantum hamiltonian contains mixing of coordinates and momenta, the naive path integral fails in general to provide the correct quantization. For a given classical hamiltonian $H(x,p)\,$,
the functional integral in phase space produces transition amplitudes corresponding to the quantum Hamiltonian $\hat H_W(\hat x,\hat p)$ obtained from the classical one by Weyl ordering \cite{Sato:1976hy}. The correctness of the choice of ${\cal H}$ as classical vertex is thus ensured by the relation \eqref{WeylOrdering}, that greatly simplifies the model.
In gauge fixing the local symmetry \eqref{Localtauinv} on the circle, it is customary to fix the einbein to a constant: $e(\tau)=T$ that plays the role of Schwinger's proper time and breaks the Weyl symmetry \eqref{ParticleWeyl}. The ghost system associated to $\tau$-reparametrizations has locally trivial action, but its Faddeev-Popov determinant contributes on $S^1$ topology with a factor of $T^{-1}\,$, yielding
\begin{equation}
\Gamma[h]=\int_0^\infty\frac{dT}{T}\int_{S^1}DxDp\,e^{-S_E[x,p;T]}\;,
\end{equation}
where the integral over $T$ is the finite dimensional remnant\footnote{The quantity $T=\int_0^1d\tau\,e(\tau)$ is gauge invariant on the circle; hence it constitutes a modulus to be integrated over after gauge fixing.} of the functional $e$-integral, and the gauge fixed action is simply obtained by replacing $e(\tau)=T\,$. We shall now compute the trace of the heat kernel
\begin{equation}
K[T;h]=\int_{S^1}DxDp\,e^{-S_E[x,p;T]}
\end{equation}
by treating the phase space vertex ${\cal H}(x,p)$ as a perturbation over the free action
\begin{equation}\label{FreeActionPX}
S_2[x,p]=\int_0^1d\tau\,\big[T\,p^2-i\,p_\mu\dot x^\mu\big]\;.
\end{equation}
First of all we shall extract the zero mode from the periodic trajectories $x^\mu(\tau)\,$:
\begin{equation}
x^\mu(\tau)=x^\mu+q^\mu(\tau)\;,\quad x^\mu:=\int_0^1d\tau\,x^\mu(\tau)\;\rightarrow\;\int_0^1d\tau\,q^\mu(\tau)=0\;,
\end{equation}
so that the functional measure splits as $\int_{S^1}Dx=\int d^dx\int_{\rm PBC'}Dq\,$, where PBC' denotes periodic boundary condition with the zero mode removed. Expectation values w.r.t. the free action \eqref{FreeActionPX} are denoted by
\begin{equation}
\left\langle F(q,p)\right\rangle:=\frac{\int_{\rm PBC'}Dq\int Dp\,F(q,p)\,e^{-S_2[q,p]}}{\int_{\rm PBC'}Dq\int Dp\,e^{-S_2[q,p]}}
\end{equation}
and the trace of the heat kernel can be written as
\begin{equation}\label{HK}
K[T;h]=\int\frac{d^d x}{(4\pi  T)^{d/2}}\,\left\langle e^{-T\int_0^1d\tau\,{\cal H}(x+q,p)}\right\rangle=\int\frac{d^d x}{(4\pi  T)^{d/2}}\sum_{n=0}^\infty T^n\,{\cal V}_n[T;h]\;,
\end{equation}
where $(4\pi T)^{-d/2}$ is the value of the free path integral and the $n$-field effective vertex is given by
\begin{equation}\label{EffectiveVertex}
\begin{split}
{\cal V}_n[T;h] &= \frac{(-1)^n}{n!}\,\int_0^1d\tau_1\cdots \int_0^1d\tau_n\,\left\langle\prod_{i=1}^n{\cal H}\big(x+q(\tau_i),p(\tau_i)\big)\right\rangle\\
&= \frac{(-1)^n}{n!}\,\int_0^1d\tau_1\cdots \int_0^1d\tau_n\,\left\langle e^{\sum_{i=1}^n q_i\cdot\partial_{x_i}+p_i\cdot\partial_{u_i}}\right\rangle\,{\cal  H}(x_1,u_1)\cdots{\cal  H}(x_n,u_n)\vert_{\substack{x_i=x\\u_i=0}}\\
&=:\hat V_n(T;\partial_{x_i},\partial_{u_i})\,{\cal  H}(x_1,u_1)\cdots{\cal  H}(x_n,u_n)\vert_{\substack{x_i=x\\u_i=0}}\;,
\end{split}
\end{equation}
where $q_i:=q(\tau_i)\,$, $p_i:=p(\tau_i)$ and we expanded the generating functions ${\cal H}(x+q_i,p_i)$ around $(x,0)\,$.
In terms of the currents
\begin{equation}\label{CurrentsPI}
j_n(\tau):=\sum_{i=1}^n\delta(\tau-\tau_i)\partial_{x_i}\;,\quad k_n(\tau):=\sum_{i=1}^n\delta(\tau-\tau_i)\partial_{u_i}\;,
\end{equation}
the quantum average above can be recast in the form of a generating functional:
\begin{equation}
\left\langle e^{\sum_{i=1}^n q_i\cdot\partial_{x_i}+p_i\cdot\partial_{u_i}}\right\rangle=\left\langle e^{\int_0^1d\tau[q(\tau)\cdot j_n(\tau)+p(\tau)\cdot k_n(\tau)]}\right\rangle\;.
\end{equation}
This is a quadratic path integral and can be computed exactly, yielding 
\begin{equation}\label{QuantumAverage}
\left\langle e^{\int_0^1d\tau[q(\tau)\cdot j_n(\tau)+p(\tau)\cdot k_n(\tau)]}\right\rangle={\rm exp}\Big\{\frac12\int_0^1d\tau\int_0^1d\sigma K_n^T(\tau)\,G(\tau,\sigma)K_n(\sigma)\Big\}
\end{equation}
for the column vector $K_n(\tau)=\left(
\begin{array}{c}
k_n(\tau)\\
j_n(\tau)\\
\end{array}
\right)\,$. Here $G(\tau,\sigma)$ is the matrix of the phase space Green's functions\footnote{See Appendix \ref{PSprops} for details.}
\begin{equation}
\langle p_\mu(\tau)p_\nu(\sigma)\rangle=\frac{1}{2T}\,\eta_{\mu\nu}\;,\quad \langle p_\mu(\tau)q^\nu(\sigma)\rangle=i\,\delta^\nu_\mu\,f(\sigma-\tau)\;,\quad \langle q^\mu(\tau)q^\nu(\sigma)\rangle=2T\,\eta^{\mu\nu}\,g(\tau-\sigma)
\end{equation}
with propagators
\begin{equation}
f(\tau)=-\tau+\tfrac12\,{\rm sign}(\tau)\;,\quad g(\tau)=\tfrac12\,(\tau^2-\lvert\tau\rvert+\tfrac16)\;,\quad \tau\in[-1,1]\;.
\end{equation}
By using the currents \eqref{CurrentsPI} in \eqref{QuantumAverage} one obtains
\begin{equation}
\begin{split}
\hat V_n(T;\partial_{x_i},\partial_{u_i}) &= \frac{(-1)^n}{n!}\int_0^1d\tau_1\cdots\int_0^1d\tau_n\\
\times&{\rm exp}\,\frac12\sum_{i,j=1}^n\left[\frac{1}{2T}\,\partial_{u_i}\cdot\partial_{u_j}+2i\,f(\tau_i-\tau_j)\,\partial_{x_i}\cdot\partial_{u_j}+2T\,g(\tau_i-\tau_j)\,\partial_{x_i}\cdot\partial_{x_j}\right]\;.
\end{split}
\end{equation}
To manipulate it further, let us notice that $\sum_{i=1}^n\partial_{x_i}\sim0$ is a total derivative, according to \eqref{EffectiveVertex} and \eqref{HK}. This allows to consistently drop the constant part in every $g(\tau)$ propagator, leaving the effective propagator $\hat g(\tau):=\frac12(\tau^2-\lvert\tau\rvert)\,$. The rigid translation invariance under $\tau_i\rightarrow\tau_i+c\,$, together with the periodicity of the trajectories over $S^1\,$, allows to fix one $\tau$ variable\footnote{This can be seen by just changing variables in integrals of periodic and translation invariant functions. However, a more precise justification comes from the gauge fixing procedure on the circle: The einbein $e(\tau)$ possesses, on $S^1$ topology, a Killing vector that is not fixed by the gauge $e(\tau)=T$ and that generates global translations around the circle. A natural way to fix the leftover global symmetry  is then to fix the position of one vertex on the circle, \emph{e.g.} by setting $\tau_n=0\,$, as it is customary in String Theory.}, let us say $\tau_n=0$ and, thanks to the symmetry under permutations of the  $\tau_i\,$, that is manifest from the first line of \eqref{EffectiveVertex}, one can also transform the $\tau$-integral: $\int_0^1d\tau_1...\int_0^1d\tau_{n-1}\rightarrow(n-1)!\int_0^1d\tau_1\int_0^{\tau_1}d\tau_2...\int_0^{\tau_{n-2}}d\tau_{n-1}\,$, finally obtaining for the effective vertex
\begin{equation}\label{masterformula}
\begin{split}
&{\cal V}_n[T;h]=\frac{(-1)^n}{n}e^{\frac{1}{4T}\,\partial_U^2}\int_0^1d\tau_1\int_0^{\tau_1}d\tau_2\cdots\int_0^{\tau_{n-2}}d\tau_{n-1}\\
&\times\,{\rm exp}\,\sum_{i<j}^{(\tau_n=0)}\left\{-i(\tau_{ij}-\tfrac12)(\partial_{x_i}\cdot\partial_{u_j}-\partial_{x_j}\cdot\partial_{u_i})+T\,\tau_{ij}(\tau_{ij}-1)\partial_{x_i}\cdot\partial_{x_j}\right\}\,{\cal  H}(x_1,u_1)\cdots{\cal  H}(x_n,u_n)\vert_{\substack{x_i=x\\u_i=0}}
\end{split}
\end{equation}
where $\tau_{ij}:=\tau_i-\tau_j$ and $\partial_U:=\sum_{i=1}^n\partial_{u_i}\,$.
For any given set of spins $\{s_1,...,s_n\}$ of the CHS fields $h_{s_i}\,$, the maximal number of $u$-derivatives is bounded by $S:=\sum_{i=1}^ns_i\,$, so that the exponential $e^{\partial_U^2/4T}$ contributes with the maximal negative power $T^{-[S/2]}\,$, making the Laurent expansion
\begin{equation}
{\cal V}_n[T;h]=\sum_{k=-\infty}^\infty T^k\,{\cal V}^{(k)}_n[h]
\end{equation}
well defined. The coefficient giving rise to the logarithmic divergence is thus the one of order ${k=\tfrac{d}{2}-n}\,$, and the CHS action can be identified as\footnote{The field independent ${\cal V}_0(T)=1$ cannot contribute to the logarithmic divergence and neither can the linear ${\cal V}_1[T;h]$ in $d\geq4\,$.}
\begin{equation}\label{CHSaction}
S_{\rm CHS}[h]=\int d^dx \sum_{n=2}^\infty{\cal V}^{(d/2-n)}_n[h]\;.
\end{equation}
All the vertices of \eqref{CHSaction} can be in principle computed using \eqref{masterformula} but, since locality has to be manifest in the spin decomposition of the $h(x,u)$ basis, it would be desirable to develop a formalism that avoids the introduction of the ${\cal H}(x,u)$ generating function. In such a case, all the differential operators $\hat{ V}_n(T;\partial_{x_i},\partial_{u_i})$ should reduce to finite polynomials of homogeneous degree in spacetime derivatives. 

\subsection{The quadratic action}
The effective vertex \eqref{masterformula} is an equivalent representation of the one obtained in \cite{Bekaert:2010ky}, but for illustrative purposes we shall rederive the quadratic action by computing  
${\cal V}^{(\frac{d}{2}-2)}_2[h]\,$. From \eqref{masterformula} one has
\begin{equation}
{\cal V}_2[T;h]=\tfrac12\,e^{\frac{1}{4T}\partial_U^2+\frac{i}{2}\partial_x\cdot\partial_U}{\cal F}(-i\partial_x\cdot\partial_U,T\,\Box)\,{\cal H}(x,u){\cal H}(x',u')\vert_{\substack{x'=x\\u,u'=0}}
\end{equation}
where we used $\partial_{x'}\sim-\partial_x$ and
\begin{equation}
{\cal F}(\alpha,\beta):=\int_0^1d\tau\,e^{\alpha\tau+\beta\tau(1-\tau)}=\sum_{n=0}^\infty\frac{1}{n!}\,{\cal F}_n(\alpha)\,\beta^n\;.
\end{equation}
The functions ${\cal F}_n(\alpha)$ can be computed as hypergeometric integrals \eqref{HyperInt}: 
\begin{equation}
{\cal F}_n(\alpha)=\frac{(n!)^2}{(2n+1)!}\,{}_1F_1(n+1;2n+2;\alpha)\;,
\end{equation}
and using Kummer's formula \eqref{Kummer} can be recast in terms of Bessel functions \eqref{Bessel}, giving
\begin{equation}
{\cal V}_2[T;h]=\tfrac{\sqrt{\pi}}{2}\,e^{\frac{1}{4T}\partial_U^2}\sum_{n=0}^\infty\left(\partial_x\cdot\partial_U\right)^{-n-\frac12}J_{n+\frac12}\left(\frac{\partial_x\cdot\partial_U}{2}\right)\,(T\,\Box)^n\,{\cal H}(x,u){\cal H}(x',u')\vert_{\substack{x'=x\\u,u'=0}}\;.
\end{equation}
It is now possible to extract the contribution of order $T^{\frac{d}{2}-2}\,$, yielding
\begin{equation}
{\cal V}_2^{\left(\frac{d-4}{2}\right)}[h]=\sqrt{\tfrac{\pi}{8}}\Big[\tfrac12\sqrt{-\partial_{U\perp}^2\partial_x^2}\Big]^{-\frac{d-3}{2}}J_{\frac{d-3}{2}}\Big(\tfrac12\sqrt{-\partial_{U\perp}^2\partial_x^2}\Big)\,\left(\frac{\Box}{2}\right)^{\frac{d-4}{2}}{\cal H}(x,u){\cal H}(x',u')\vert_{\substack{x'=x\\u,u'=0}}\;,
\end{equation}
where we used Lommel's expansion formula \eqref{Lommel}, and the transverse projection is defined as 
\begin{equation}
v^\mu_\perp:=v^\mu-\frac{v\cdot\partial_x\partial_x^\mu}{\Box}\;,
\end{equation}
so that $\partial_{U\perp}^2\Box=\partial_U^2\Box-(\partial_x\cdot\partial_U)^2\,$. The above result coincides with the one of \cite{Bekaert:2010ky}, and it can be seen that the Bessel function ``undresses'' the ${\cal H}$ fields, leaving a finite degree polynomial acting on the two $h$ fields. To do so one applies Gegenbauer addition theorem \eqref{GegenbauerAdd} to the above Bessel function, with the triplet of variables $Z:=\tfrac12\sqrt{-\partial_{U\perp}^2\Box}$ and\footnote{In the variables $z_i$ one can exchange $x$ with $x'$ for free.} $z_i:=\tfrac12\sqrt{-\partial_{u_i\perp}^2\Box_i}$ obeying 
\begin{equation}
Z^2=z^2+z'^2-2\,zz'\,w\quad{\rm for}\quad w=\frac{\partial_{u\perp}\cdot\partial_{u'\perp}}{\sqrt{\partial_{u\perp}^2\partial_{u'\perp}^2}}\;.
\end{equation}
The resulting $J_{n+\frac{d-3}{2}}(z_i)$ that appear from the addition theorem produce the inverse maps ${\cal P}^{-1}_d$ when acting on the corresponding ${\cal H}(x_i,u_i)$ as\footnote{For details see the original derivation \cite{Bekaert:2010ky}.} 
\begin{equation}
\left.\left[z_i^{-n-\frac{d-3}{2}}J_{n+\frac{d-3}{2}}(z_i)\,{\cal  H}(x_i,u_i)\right]\right\rvert_n=h_n(x_i,u_i)\;,
\end{equation}
and one is left with
\begin{equation}
{\cal V}_2^{\left(\frac{d-4}{2}\right)}[h]=\sqrt{\tfrac{\pi}{8}}\sum_{n=0}^\infty c_n(d)\,\left(\frac{\Box}{2}\right)^{n+\frac{d-4}{2}}\left(\partial_{u\perp}^2\partial_{u'\perp}^2\right)^{n/2}C_n^{\frac{d-3}{2}}\left(\frac{\partial_{u\perp}\cdot\partial_{u'\perp}}{\sqrt{\partial_{u\perp}^2\partial_{u'\perp}^2}}\right)\,h(x,u)h(x',u')\vert_{\substack{x'=x\\u,u'=0}}\;,
\end{equation}
where $C_n^\nu(w)$ is the Gegenbauer polynomial and $c_n(d):=\frac{2^{-3n-\frac{d-3}{2}}}{\Gamma(n+\frac{d-1}{2})(\frac{d-3}{2})_n}\,$. From the form of $\partial_{u_i\perp}$ one can see that the above expression is indeed local for each $n$ and of homogeneous degree ${2n+d-4}$ in spacetime derivatives. It is also easy to view, from the definition of Gegenbauer polynomials \eqref{Gegenbauer}, that the sum over $n$ is diagonal in contractions, being of homogeneous degree $(\partial_u\partial_{u'})^n\,$. The above expression is proportional for each $n$ to the corresponding transverse and traceless projector\footnote{See appendix \ref{TTApp} for the explicit form of the projectors.}$P_n\,$, that can be displayed to write the quadratic action in the form \eqref{FreeCHS} that is manifestly gauge invariant under the linearized transformations \eqref{gaugelin}:
\begin{equation}
S_{2\,{\rm CHS}}=\int d^dx\,\sum_{s=0}^\infty c_s\,h_s(x,u)\,P_s\left(\overleftarrow{\partial_u},\overrightarrow{\partial_v}\right)\Box^{s+\frac{d-4}{2}}\,h_s(x,v)\;,
\end{equation}
where we discarded a spin-independent constant and $c_s=\tfrac{1}{2^{3s}\Gamma(s+\frac{d-1}{2})}\,$.

\section{Discussion and Conclusions}
\label{Discussion}

In this paper we have provided a worldline path integral representation for the non-linear conformal higher spin action \cite{Segal:2002gd,Bekaert:2010ky} in arbitrary even dimensions. We have rederived the quadratic part of the action, and we plan to come back in the future for the computation of cubic and higher vertices, some of which have been computed in \cite{Joung:2015eny,Beccaria:2016syk}, in transverse-traceless gauge, in the context of scattering amplitudes calculations.

The example of the quadratic action suggests that the ``undressing'' maps ${\cal P}_d^{-1}$ should appear at all orders in the differential operators $\hat V_n(T;\partial_{x_i},\partial_{u_i})\,$, leaving finite degree polynomials acting on a string of fields $h_{s_1}...h_{s_n}\,$. From the representation \eqref{masterformula} it is not transparent how this should take place. For this reason, it would be interesting to find a way to avoid the introduction of the dressed generating function ${\cal H}(x,u)\,$, and work directly in the basis of CHS $h(x,u)\,$. 
To this goal, when restricting to four dimensions, a considerable advantage could come by working in terms of $sl(2,\mathbb{C})$ spinors instead of tensors. All the trace projections would become trivial and one could work directly in terms of conformal primary currents $J_s\,$.

Another issue of (non-linear) field redefinitions is apparent when looking at the low spin content of the Noether interaction \eqref{Noether}: the linear coupling
\begin{equation}
\sum_{s=0}^2h_sJ_s\sim h_0\,\phi^*\phi+h_1^\mu\,\phi^*\partial_\mu\phi+\tfrac12\,h_2^{\mu\nu}\,\phi^*\partial_\mu\partial_\nu\phi
\end{equation}
does not coincide with the standard Weyl and $U(1)$ invariant coupling of a complex scalar to a vector gauge field in curved spacetime, \emph{i.e.}
\begin{equation}
S=\int d^dx\sqrt{g}\,\Big[g^{\mu\nu}D_\mu\phi^*D_\nu\phi+\tfrac{d-2}{4(d-1)}\,R\,\phi^*\phi\Big]\;,
\end{equation}
and the basis $(h_0,h_1,h_2)$ is related to the geometric ${(A_\mu,\,g_{\mu\nu}=\eta_{\mu\nu}+h_{\mu\nu})}$ by a non-linear field redefinition. This issue has been discussed, for instance, in \cite{Segal:2002gd,Bekaert:2010ky,Beccaria:2016syk}. In fact, the covariant description of CHS fields in curved (maybe Bach-flat) background is still an open problem \cite{Nutma:2014pua,Grigoriev:2016bzl,Beccaria:2017nco}, and it underpins the question of vanishing Weyl anomalies. To this end, it would be interesting to find a first quantized origin of CHS fields, since at the worldline (or worldsheet) level it could be easier to achieve a covariant description, and the sum over spins, that is crucial in proving the vanishing of anomalies as well as triviality of scattering amplitudes, would be accounted for by the worldline fields.

\vspace{6pt}

\section*{Acknowledgments}
We would like to thank Arkady Tseytlin, Evgeny Skvortsov, Dmitry Ponomarev and David De Filippi for useful discussions. The author thanks the Imperial College London and the Ludwig Maximilian University of Munich for kind hospitality during the final stages of this work. The work of R.B. was supported by a PDR “Gravity and
extensions” from the F.R.S.-FNRS (Belgium).

\appendix

\section{Worldline phase space propagators}
\label{PSprops}

In this section we will derive the phase space propagators associated with the euclidean worldline action on the circle
\begin{equation}\label{S2}
S_2[x,p]=\int_0^1d\tau\,\big[T\,p^2-i\,p_\mu\dot x^\mu\big]\;.
\end{equation}
Upon extracting the zero mode from the periodic trajectories $x^\mu(\tau)\,$, one goes to Fourier space
\begin{equation}
\begin{split}
&x^\mu(\tau)=x_0^\mu+q^\mu(\tau)\;,\quad q^\mu(\tau)=\sum_{n\in\mathbb{Z}\setminus\{0\}}q^\mu_n\,e^{2\pi in\tau}\;,\\ 
&p_\mu(\tau)=\sum_{n\in\mathbb{Z}}p_{\mu\,n}\,e^{2\pi in\tau}\;.
\end{split}
\end{equation}
The Fourier modes obey the reality conditions
\begin{equation}
(x_0,p_0)\in\mathbb{R}\;,\quad q_n^*=q_{-n}\;,\quad p_n^*=p_{-n}
\end{equation}
and the action \eqref{S2} reads
\begin{equation}
S_2=T\,p_0^2+\sum_{n=1}^\infty Z_n^\dagger K_n\,Z_n
\end{equation}
in terms of the phase space vector $Z_n$ and kinetic matrix $K_n$
\begin{equation}
Z_n:=\begin{pmatrix}p_n\\q_n
\end{pmatrix}\;,\quad K_n:=\begin{pmatrix}2T & 2\pi n\\[2mm]-2\pi n & 0
\end{pmatrix}\;,
\end{equation}
where we suppressed all the spacetime indices. From the inverse matrix
\begin{equation}
K_n^{-1}=\begin{pmatrix}0 & -\tfrac{1}{2\pi n}\\[2mm]\tfrac{1}{2\pi n} & \tfrac{2T}{4\pi^2 n^2}
\end{pmatrix}
\end{equation}
it is immediate to extract the two-point functions
\begin{equation}
\left\langle p_\mu(\tau)\,p_\nu(\sigma)\right\rangle=\frac{\eta_{\mu\nu}}{2T}\;,\quad \left\langle p_\mu(\tau)\,q^\nu(\sigma)\right\rangle=i\,\delta_\mu^\nu\,f(\sigma-\tau)\;,\quad \left\langle q^\mu(\tau)\,q^\nu(\sigma)\right\rangle=2T\,\eta_{\mu\nu}\,g(\tau-\sigma)\;.
\end{equation}
The above propagators are defined in terms of their Fourier series and read
\begin{equation}
\begin{split}
f(\tau) &:= \sum_{n=1}^\infty\frac{1}{\pi n}\,\sin(2\pi n\,\tau)=-\tau+\tfrac12\,{\rm sign}(\tau)\;,\\
g(\tau)&:= \sum_{n=1}^\infty\frac{1}{2\pi^2 n^2}\,\cos(2\pi n\,\tau)=\tfrac12\,\tau^2-\tfrac12\,\lvert\tau\rvert+\tfrac{1}{12}\;,
\end{split}
\end{equation}
where the latter expressions in terms of elementary functions hold in the interval $[-1,1]\,$.

\section{Special functions}

We collect here the definitions and formulas that are relevant to the main text. The Bessel function of the first kind can be defined by the series
\begin{equation}\label{Bessel}
J_\nu(z):=\sum_{k=0}^\infty\frac{(-1)^k}{k!\,\Gamma(\nu+k+1)}\,\left(\frac{z}{2}\right)^{\nu+2k}\;,
\end{equation}
while the modified Bessel function $I_\nu(z)$ is given by
\begin{equation}
I_\nu(z):=\sum_{k=0}^\infty\frac{1}{k!\,\Gamma(\nu+k+1)}\,\left(\frac{z}{2}\right)^{\nu+2k}=i^{-\nu}J_\nu(iz)\;.
\end{equation}
The Lommel expansion formula reads
\begin{equation}\label{Lommel}
\sqrt{z+h}^{-\nu}J_\nu\left(\sqrt{z+h}\right)=\sum_{k=0}^\infty\tfrac{1}{k!}\left(-\tfrac{h}{2}\right)^k\sqrt{z}^{-\nu-k}J_{\nu+k}\left(\sqrt{z}\right)
\end{equation}
and, for a triplet $(\omega,x,y)$ obeying
\begin{equation}
\omega^2=x^2+y^2-2\,xy\,\cos\theta\;,
\end{equation}
one has the Gegenbauer addition theorem:
\begin{equation}\label{GegenbauerAdd}
\omega^{-\nu}J_\nu(\omega)=2^\nu\Gamma(\nu)\sum_{n=0}^\infty(\nu+n)\,x^{-\nu}J_{\nu+n}(x)\,y^{-\nu}J_{\nu+n}(y)\,C_n^\nu(\cos\theta)\;,
\end{equation}
where $C_n^\nu(z)$ is the Gegenbauer polynomial defined by
\begin{equation}\label{Gegenbauer}
C_n^\nu(z)=\sum_{k=0}^{[n/2]}\frac{(-1)^k(\nu)_{n-k}}{k!\,(n-2k)!}\,(2z)^{n-2k}\;.
\end{equation}
The generalized hypergeometric function ${}_pF_q$ is defined by the series
\begin{equation}
{}_pF_q(a_1,...,a_p;b_1,...,b_q;z)=\sum_{n=0}^\infty\frac{(a_1)_n\cdots(a_p)_n}{(b_1)_n\cdots(b_q)_n}\frac{z^n}{n!}\;.
\end{equation}
The confluent hypergeometric function ${}_1F_1(a;b;z)$ admits the integral representation
\begin{equation}\label{HyperInt}
{}_1F_1(a;b;z)=\frac{\Gamma(b)}{\Gamma(a)\Gamma(b-a)}\int_0^1du\,e^{zu}u^{a-1}(1-u)^{b-a-1}
\end{equation}
and for $b=2a$ it is related to the ${}_0F_1$ series and thus to the Bessel function via Kummer's formula:
\begin{equation}\label{Kummer}
{}_1F_1(a;2a;z)=e^{z/2}{}_0F_1\left(;a+\tfrac12;\tfrac{z^2}{16}\right)=e^{z/2}\left(\frac{z}{4}\right)^{\frac12-a}\Gamma(a+\tfrac12)\,I_{a-\frac12}(z/2)\;.
\end{equation}

\section{Transverse-traceless projectors}
\label{TTApp}

The transverse-traceless projectors $P_{\mu(s)}{}^{\nu(s)}$ of spin $s\,$, obeying
\begin{equation}
\eta^{\alpha\beta}\,P_{\alpha\beta\mu(s-2)}{}^{\nu(s)}=0=P_{\mu(s)}{}^{\alpha\beta\nu(s-2)}\,\eta_{\alpha\beta}\;,\quad \partial^\alpha P_{\alpha\mu(s-1)}{}^{\nu(s)}=0=P_{\mu(s)}{}^{\alpha\nu(s-1)}\partial_\alpha
\end{equation}
with normalization
\begin{equation}
P_{\mu(s)}{}^{\lambda(s)}P_{\lambda(s)}{}^{\nu(s)}=P_{\mu(s)}{}^{\nu(s)}\;,
\end{equation}
can be built from $s$ powers of the corresponding spin one transverse projector
\begin{equation}
P_\mu{}^\nu:=\delta_\mu^\nu-\frac{\partial_\mu\partial^\nu}{\Box}
\end{equation}
as
\begin{equation}
P_{\mu(s)}{}^{\nu(s)}=\sum_{k=0}^{[s/2]}\alpha_k(s)\left(P_{\mu\mu}\,P^{\nu\nu}\right)^k\left(P_\mu{}^\nu\right)^{s-2k}\;,
\end{equation}
with coefficients $\alpha_k(s)$ being fixed by tracelessness 
as
\begin{equation}
\alpha_k(s)=\frac{(-1)^ks!\,\Gamma\big(s-k+\frac{d-3}{2}\big)}{4^k\,k!(s-2k)!\,\Gamma\big(s+\frac{d-3}{2}\big)}\;.
\end{equation}
The generating function of the spin $s$ projector:
\begin{equation}
P_s(u,v):=\frac{1}{s!}\,u^{\mu_1}...\,u^{\mu_s}\,P_{\mu(s)}{}^{\nu(s)}\,v_{\nu_1}...\,v_{\nu_s}\;,
\end{equation}
acts on the generating function of a spin $s$ field $h_s(x,u)=\tfrac{1}{s!}\,h_{\mu(s)}(u^\mu)^s$ as
\begin{equation}
\big(P_s\,h_s\big)(x,u):=\tfrac{1}{s!}\,P_{\mu(s)}{}^{\nu(s)}h_{\nu(s)}(x)(u^\mu)^s=P_s(u,\partial_v)\,h_s(x,v)\;,
\end{equation}
and it can be written in terms of Gegenbauer polynomials as
\begin{equation}\label{TTprojector}
P_s(u,v)=\frac{\Gamma\big(\frac{d-3}{2}\big)}{2^s\,\Gamma\big(s+\frac{d-3}{2}\big)}\,\left(\lvert u_\perp\rvert\lvert v_\perp\rvert\right)^s\,C_s^{\frac{d-3}{2}}\left(\frac{u_\perp\cdot v_\perp}{\lvert u_\perp\rvert\lvert v_\perp\rvert}\right)\;,
\end{equation}
where transverse vectors are defined by
\begin{equation}
u^\mu_\perp:=u^\mu-\frac{u\cdot\partial\partial^\mu}{\Box}\;.
\end{equation}

\bibliographystyle{mdpi}

\begin{thebibliography}{999}

\bibitem{Kaku:1978nz}
  M.~Kaku, P.~K.~Townsend and P.~van Nieuwenhuizen,
  Properties of Conformal Supergravity,
  {\em Phys.\ Rev.}\ D {\bf 17} (1978) 3179.

\bibitem{Bergshoeff:1980is}
  E.~Bergshoeff, M.~de Roo and B.~de Wit,
  Extended Conformal Supergravity,
  {\em Nucl.\ Phys.}\ B {\bf 182} (1981) 173.
  
\bibitem{Fradkin:1985am}
  E.~S.~Fradkin and A.~A.~Tseytlin,
  Conformal Supergravity,
  {\em Phys.\ Rept.}\  {\bf 119} (1985) 233.
  
\bibitem{Siegel:1988gd}
W.~Siegel, All Free Conformal Representations in All Dimensions,  {\em
  Int. J. Mod. Phys.} {\bf A4} (1989) 2015.

\bibitem{Fradkin:1989md}
  E.~S.~Fradkin and V.~Y.~Linetsky,
  Cubic Interaction in Conformal Theory of Integer Higher Spin Fields in Four-dimensional Space-time,
  {\em Phys.\ Lett.}\ B {\bf 231} (1989) 97.


\bibitem{Tseytlin:2002gz}
  A.~A.~Tseytlin,
  On limits of superstring in $AdS_5 \times S^5$,
  {\em Theor.\ Math.\ Phys.}\  {\bf 133} (2002) 1376,
  [\href{http://xxx.lanl.gov/abs/hep-th/0201112}{{\tt
  hep-th/0201112}}].

\bibitem{Segal:2002gd}
A.~Y. Segal, Conformal higher spin theory,  {\em Nucl. Phys.} {\bf
  B664} (2003) 59--130, [\href{http://xxx.lanl.gov/abs/hep-th/0207212}{{\tt
  hep-th/0207212}}].

\bibitem{Shaynkman:2004vu}
O.~V. Shaynkman, I.~{\relax Yu}. Tipunin, and M.~A. Vasiliev, Unfolded
  form of conformal equations in M dimensions and o(M + 2) modules,  {\em
  Rev. Math. Phys.} {\bf 18} (2006) 823--886,
  [\href{http://xxx.lanl.gov/abs/hep-th/0401086}{{\tt hep-th/0401086}}].

\bibitem{Marnelius:2008er}
R.~Marnelius, Lagrangian conformal higher spin theory,
  \href{http://xxx.lanl.gov/abs/0805.4686}{{\tt arXiv:0805.4686}}.

\bibitem{Vasiliev:2009ck}
M.~A. Vasiliev, Bosonic conformal higher-spin fields of any symmetry,
  {\em Nucl. Phys.} {\bf B829} (2010) 176--224,
  [\href{http://xxx.lanl.gov/abs/0909.5226}{{\tt arXiv:0909.5226}}].
  
\bibitem{Bekaert:2010ky}
  X.~Bekaert, E.~Joung and J.~Mourad,
  Effective action in a higher-spin background,
  {\em JHEP} {\bf 1102} (2011) 048,
  [\href{http://xxx.lanl.gov/abs/1012.2103}{{\tt arXiv:1012.2103}}].

\bibitem{Bandos:2011wi}
I.~A. Bandos, J.~A. de~Azcarraga, and C.~Meliveo, Extended supersymmetry in massless conformal higher spin theory,  {\em Nucl. Phys.} {\bf B853}
  (2011) 760--776, [\href{http://xxx.lanl.gov/abs/1106.5199}{{\tt
  arXiv:1106.5199}}].

\bibitem{Bekaert:2012vt}
X.~Bekaert and M.~Grigoriev, Notes on the ambient approach to boundary
  values of AdS gauge fields,  {\em J.~Phys.} {\bf A46} (2013) 214008,
  [\href{http://xxx.lanl.gov/abs/1207.3439}{{\tt arXiv:1207.3439}}].

\bibitem{Haehnel:2016mlb}
P.~Haehnel and T.~McLoughlin, Conformal Higher Spin Theory and Twistor Space Actions,  [\href{http://xxx.lanl.gov/abs/1604.08209}{{\tt
  arXiv:1604.08209}}].
  
\bibitem{Kuzenko:2016qdw}
  S.~M.~Kuzenko,
  Higher spin super-Cotton tensors and generalisations of the linear–chiral duality in three dimensions,
  {\em Phys.\ Lett.}\ B {\bf 763} (2016) 308
 [\href{http://xxx.lanl.gov/abs/1606.08624}{{\tt
  arXiv:1606.08624}}].
  
\bibitem{Kuzenko:2017ujh}
  S.~M.~Kuzenko, R.~Manvelyan and S.~Theisen,
 Off-shell superconformal higher spin multiplets in four dimensions,
  {\em JHEP} {\bf 1707} (2017) 034
 [\href{http://xxx.lanl.gov/abs/1701.00682}{{\tt
  arXiv:1701.00682}}].
  
\bibitem{Metsaev:2007fq}
  R.~R.~Metsaev,
  Ordinary-derivative formulation of conformal low spin fields,
  {\em JHEP} {\bf 1201} (2012) 064,
 [\href{http://xxx.lanl.gov/abs/0707.4437}{{\tt
  arXiv:0707.4437}}].
  
\bibitem{Metsaev:2007rw}
  R.~R.~Metsaev,
  Ordinary-derivative formulation of conformal totally symmetric arbitrary spin bosonic fields,
  {\em JHEP} {\bf 1206} (2012) 062,
 [\href{http://xxx.lanl.gov/abs/0709.4392}{{\tt
  arXiv:0709.4392}}].
  
\bibitem{Nutma:2014pua}
  T.~Nutma and M.~Taronna,
  On conformal higher spin wave operators,
  {\em JHEP} {\bf 1406} (2014) 066,
 [\href{http://xxx.lanl.gov/abs/1404.7452}{{\tt
  arXiv:1404.7452}}].

\bibitem{Grigoriev:2016bzl}
  M.~Grigoriev and A.~A.~Tseytlin,
  On conformal higher spins in curved background,
  {\em J.\ Phys.}\ A {\bf 50} (2017) no.12,  125401,
   [\href{http://xxx.lanl.gov/abs/1609.09381}{{\tt
  arXiv:1609.09381}}].
  
\bibitem{Beccaria:2017nco}
  M.~Beccaria and A.~A.~Tseytlin,
  On induced action for conformal higher spins in curved background,
  {\em Nucl.\ Phys.}\ B {\bf 919} (2017) 359,
   [\href{http://xxx.lanl.gov/abs/1702.00222}{{\tt
  arXiv:1702.00222}}].
  
\bibitem{Fradkin:1981jc}
  E.~S.~Fradkin and A.~A.~Tseytlin,
  One Loop Beta Function in Conformal Supergravities,
  {\em Nucl.\ Phys.}\ B {\bf 203} (1982) 157.
  
\bibitem{Fradkin:1983tg}
  E.~S.~Fradkin and A.~A.~Tseytlin,
  Conformal Anomaly in Weyl Theory and Anomaly Free Superconformal Theories,
  {\em Phys.\ Lett.}\  {\bf 134B} (1984) 187.
  
\bibitem{Giombi:2013yva}
  S.~Giombi, I.~R.~Klebanov, S.~S.~Pufu, B.~R.~Safdi and G.~Tarnopolsky,
  AdS Description of Induced Higher-Spin Gauge Theory,
  {\em JHEP} {\bf 1310} (2013) 016,
[\href{http://xxx.lanl.gov/abs/1306.5242}{{\tt
  arXiv:1306.5242}}].
   
  
\bibitem{Tseytlin:2013jya}
  A.~A.~Tseytlin,
  On partition function and Weyl anomaly of conformal higher spin fields,
  {\em Nucl.\ Phys.}\ B {\bf 877} (2013) 598,
   [\href{http://xxx.lanl.gov/abs/1309.0785}{{\tt
  arXiv:1309.0785}}].

\bibitem{Giombi:2014iua}
  S.~Giombi, I.~R.~Klebanov and B.~R.~Safdi,
  Higher Spin AdS$_{d+1}$/CFT$_d$ at One Loop,
  {\em Phys.\ Rev.}\ D {\bf 89} (2014) no.8,  084004,
   [\href{http://xxx.lanl.gov/abs/1401.0825}{{\tt
  arXiv:1401.0825}}].

\bibitem{Beccaria:2014xda}
  M.~Beccaria and A.~A.~Tseytlin,
  Higher spins in AdS$_{5}$ at one loop: vacuum energy, boundary conformal anomalies and AdS/CFT,
  {\em JHEP} {\bf 1411} (2014) 114,
[\href{http://xxx.lanl.gov/abs/1410.3273}{{\tt
  arXiv:1410.3273}}].
  
\bibitem{Beccaria:2015vaa}
  M.~Beccaria and A.~A.~Tseytlin,
 On higher spin partition functions,
 {\em J.\ Phys.}\ A {\bf 48} (2015) no.27,  275401,
[\href{http://xxx.lanl.gov/abs/1503.08143}{{\tt
  arXiv:1503.08143}}].
  
\bibitem{Beccaria:2017lcz}
  M.~Beccaria and A.~A.~Tseytlin,
  C$_{T}$ for conformal higher spin fields from partition function on conically deformed sphere,
  {\em JHEP} {\bf 1709} (2017) 123
 [\href{http://xxx.lanl.gov/abs/1707.02456}{{\tt
  arXiv:1707.02456}}].
  
\bibitem{Vasiliev:1990en}
  M.~A.~Vasiliev,
  Consistent equation for interacting gauge fields of all spins in (3+1)-dimensions,
  {\em Phys.\ Lett.}\ B {\bf 243} (1990) 378.
  
\bibitem{Vasiliev:1990vu}
  M.~A.~Vasiliev,
  Properties of equations of motion of interacting gauge fields of all spins in (3+1)-dimensions,
  {\em Class.\ Quant.\ Grav.}\  {\bf 8} (1991) 1387.

\bibitem{Vasiliev:1992av}
  M.~A.~Vasiliev,
  More on equations of motion for interacting massless fields of all spins in (3+1)-dimensions,
  {\em Phys.\ Lett.}\ B {\bf 285} (1992) 225.
  
\bibitem{Vasiliev:1999ba}
  M.~A.~Vasiliev,
  Higher spin gauge theories: Star product and AdS space,
  In *Shifman, M.A. (ed.): The many faces of the superworld* 533-610,
  [\href{http://xxx.lanl.gov/abs/hep-th/9910096}{{\tt
  hep-th/9910096}}].
  

\bibitem{Vasiliev:2003ev}
  M.~A.~Vasiliev,
  Nonlinear equations for symmetric massless higher spin fields in (A)dS(d),
  {\em Phys.\ Lett.}\ B {\bf 567} (2003) 139
    [\href{http://xxx.lanl.gov/abs/hep-th/0304049}{{\tt
  hep-th/0304049}}].
  
\bibitem{Bekaert:2005vh}
  X.~Bekaert, S.~Cnockaert, C.~Iazeolla and M.~A.~Vasiliev,
  Nonlinear higher spin theories in various dimensions,
  [\href{http://xxx.lanl.gov/abs/hep-th/0503128}{{\tt
  hep-th/0503128}}].
  
\bibitem{Didenko:2014dwa}
  V.~E.~Didenko and E.~D.~Skvortsov,
  Elements of Vasiliev theory,
[\href{http://xxx.lanl.gov/abs/1401.2975}{{\tt
  arXiv:1401.2975}}].  

\bibitem{Sezgin:2002rt}
  E.~Sezgin and P.~Sundell,
  Massless higher spins and holography,
  {\em Nucl.\ Phys.}\ B {\bf 644} (2002) 303
   Erratum: [{\em Nucl.\ Phys.}\ B {\bf 660} (2003) 403],
[\href{http://xxx.lanl.gov/abs/hep-th/0205131}{{\tt
  hep-th/0205131}}].  
 

\bibitem{Klebanov:2002ja}
  I.~R.~Klebanov and A.~M.~Polyakov,
 AdS dual of the critical O(N) vector model,
 {\em Phys.\ Lett.}\ B {\bf 550} (2002) 213,
[\href{http://xxx.lanl.gov/abs/hep-th/0210114}{{\tt
  hep-th/0210114}}].  
  
\bibitem{Giombi:2009wh}
  S.~Giombi and X.~Yin,
  Higher Spin Gauge Theory and Holography: The Three-Point Functions,
  {\em JHEP} {\bf 1009} (2010) 115,
 [\href{http://xxx.lanl.gov/abs/0912.3462}{{\tt
  arXiv:0912.3462}}].

\bibitem{Giombi:2010vg}
  S.~Giombi and X.~Yin,
 Higher Spins in AdS and Twistorial Holography,
  {\em JHEP} {\bf 1104} (2011) 086,
   [\href{http://xxx.lanl.gov/abs/1004.3736}{{\tt
  arXiv:1004.3736}}].

\bibitem{Giombi:2012ms}
  S.~Giombi and X.~Yin,
 The Higher Spin/Vector Model Duality,
  {\em J.\ Phys.}\ A {\bf 46} (2013) 214003,
   [\href{http://xxx.lanl.gov/abs/1208.4036}{{\tt
  arXiv:1208.4036}}].
  
\bibitem{Giombi:2016ejx}
  S.~Giombi,
  Higher Spin — CFT Duality,
\href{http://xxx.lanl.gov/abs/1607.02967}{{\tt
  arXiv:1607.02967}}.

\bibitem{Liu:1998bu}
  H.~Liu and A.~A.~Tseytlin,
  D = 4 superYang-Mills, D = 5 gauged supergravity, and D = 4 conformal supergravity,
  {\em Nucl.\ Phys.}\ B {\bf 533} (1998) 88,
 [\href{http://xxx.lanl.gov/abs/hep-th/9804083}{{\tt
  hep-th/9804083}}].
    
\bibitem{Boulanger:2011dd}
  N.~Boulanger and P.~Sundell,
  An action principle for Vasiliev's four-dimensional higher-spin gravity,
  {\em J.\ Phys.}\ A {\bf 44} (2011) 495402,
  [\href{http://xxx.lanl.gov/abs/1102.2219}{{\tt
  arXiv:1102.2219}}].
  
\bibitem{Boulanger:2012bj}
  N.~Boulanger, N.~Colombo and P.~Sundell,
 A minimal BV action for Vasiliev's four-dimensional higher spin gravity,
  {\em JHEP} {\bf 1210} (2012) 043,
  [\href{http://xxx.lanl.gov/abs/1205.3339}{{\tt
  arXiv:1205.3339}}].

\bibitem{Boulanger:2015kfa}
  N.~Boulanger, E.~Sezgin and P.~Sundell,
  4D Higher Spin Gravity with Dynamical Two-Form as a Frobenius-Chern-Simons Gauge Theory,
  [\href{http://xxx.lanl.gov/abs/1505.04957}{{\tt
  arXiv:1505.04957}}].
  
\bibitem{Bonezzi:2015igv}
  R.~Bonezzi, N.~Boulanger, E.~Sezgin and P.~Sundell,
 An Action for Matter Coupled Higher Spin Gravity in Three Dimensions,
  {\em JHEP} {\bf 1605} (2016) 003,
  [\href{http://xxx.lanl.gov/abs/1512.02209}{{\tt
  arXiv:1512.02209}}].
  
\bibitem{Bonezzi:2016ttk}
  R.~Bonezzi, N.~Boulanger, E.~Sezgin and P.~Sundell,
 Frobenius–Chern–Simons gauge theory,
  {\em J.\ Phys.\ A} {\bf 50} (2017) no.5,  055401,
  [\href{http://xxx.lanl.gov/abs/1607.00726}{{\tt
  arXiv:1607.00726}}].
  
\bibitem{Bekaert:2014cea}
  X.~Bekaert, J.~Erdmenger, D.~Ponomarev and C.~Sleight,
  Towards holographic higher-spin interactions: Four-point functions and higher-spin exchange,
  {\em JHEP} {\bf 1503} (2015) 170,
   [\href{http://xxx.lanl.gov/abs/1412.0016}{{\tt
  arXiv:1412.0016}}].
  
\bibitem{Bekaert:2015tva}
  X.~Bekaert, J.~Erdmenger, D.~Ponomarev and C.~Sleight,
  Quartic AdS Interactions in Higher-Spin Gravity from Conformal Field Theory,
  {\em JHEP} {\bf 1511} (2015) 149,
   [\href{http://xxx.lanl.gov/abs/1508.04292}{{\tt
  arXiv:1508.04292}}].

\bibitem{Sleight:2016dba}
  C.~Sleight and M.~Taronna,
  Higher Spin Interactions from Conformal Field Theory: The Complete Cubic Couplings,
  {\em Phys.\ Rev.\ Lett.}\  {\bf 116} (2016) no.18,  181602,
     [\href{http://xxx.lanl.gov/abs/1603.00022}{{\tt
  arXiv:1603.00022}}].
  
\bibitem{Gopakumar:2003ns}
  R.~Gopakumar,
  From free fields to AdS,
 {\em Phys.\ Rev.}\ D {\bf 70} (2004) 025009,
 [\href{http://xxx.lanl.gov/abs/hep-th/0308184}{{\tt
  hep-th/0308184}}].
  
\bibitem{Gopakumar:2004qb}
  R.~Gopakumar,
  From free fields to AdS. 2.,
  {\em Phys.\ Rev.}\ D {\bf 70} (2004) 025010,
 [\href{http://xxx.lanl.gov/abs/hep-th/0402063}{{\tt
  hep-th/0402063}}]. 
  
\bibitem{Gopakumar:2005fx}
  R.~Gopakumar,
 From free fields to AdS: III,
 {\em Phys.\ Rev.}\ D {\bf 72} (2005) 066008,
  [\href{http://xxx.lanl.gov/abs/hep-th/0504229}{{\tt
  hep-th/0504229}}].

  
\bibitem{Maldacena:1997re}
  J.~M.~Maldacena,
 The Large N limit of superconformal field theories and supergravity,
  {\em Int.\ J.\ Theor.\ Phys.}\  {\bf 38} (1999) 1113
   [{\em Adv.\ Theor.\ Math.\ Phys.}\  {\bf 2} (1998) 231]
  [\href{http://xxx.lanl.gov/abs/hep-th/9711200}{{\tt
  hep-th/9711200}}].
  
\bibitem{Gubser:1998bc}
  S.~S.~Gubser, I.~R.~Klebanov and A.~M.~Polyakov,
 Gauge theory correlators from noncritical string theory,
  {\em Phys.\ Lett.}\ B {\bf 428} (1998) 105,
 [\href{http://xxx.lanl.gov/abs/hep-th/9802109}{{\tt
  hep-th/9802109}}].

\bibitem{Witten:1998qj}
  E.~Witten,
 Anti-de Sitter space and holography,
 {\em Adv.\ Theor.\ Math.\ Phys.}\  {\bf 2} (1998) 253,
   [\href{http://xxx.lanl.gov/abs/hep-th/9802150}{{\tt
  hep-th/9802150}}].
  
\bibitem{Eastwood:2002su}
  M.~G.~Eastwood,
  Higher symmetries of the Laplacian,
  {\em Annals Math.}\  {\bf 161} (2005) 1645,
     [\href{http://xxx.lanl.gov/abs/hep-th/0206233}{{\tt
  hep-th/0206233}}].

\bibitem{Balasubramanian:2000pq}
  V.~Balasubramanian, E.~G.~Gimon, D.~Minic and J.~Rahmfeld,
 Four-dimensional conformal supergravity from AdS space,
 {\em Phys.\ Rev.}\ D {\bf 63} (2001) 104009,
   [\href{http://xxx.lanl.gov/abs/hep-th/0007211}{{\tt
  hep-th/0007211}}].
 
\bibitem{Compere:2008us}
  G.~Compere and D.~Marolf,
 Setting the boundary free in AdS/CFT,
 {\em Class.\ Quant.\ Grav.}\  {\bf 25} (2008) 195014,
   [\href{http://xxx.lanl.gov/abs/0805.1902}{{\tt
  arXiv:0805.1902}}].
  
\bibitem{Schubert:2001he}
  C.~Schubert,
  Perturbative quantum field theory in the string inspired formalism,
 {\em Phys.\ Rept.}\  {\bf 355} (2001) 73,
  [\href{http://xxx.lanl.gov/abs/hep-th/0101036}{{\tt hep-th/0101036}}].
  
\bibitem{Bern:1991aq}
  Z.~Bern and D.~A.~Kosower,
  The Computation of loop amplitudes in gauge theories,
  {\em Nucl.\ Phys.}\ B {\bf 379} (1992) 451.  
  
\bibitem{Strassler:1992zr}
  M.~J.~Strassler,
 Field theory without Feynman diagrams: One loop effective actions,
 {\em Nucl.\ Phys.}\ B {\bf 385} (1992) 145,
  [\href{http://xxx.lanl.gov/abs/hep-th/9205205}{{\tt
  hep-th/9205205}}].

\bibitem{Bastianelli:1992ct}
  F.~Bastianelli and P.~van Nieuwenhuizen,
  Trace anomalies from quantum mechanics,
 {\em Nucl.\ Phys.}\ B {\bf 389} (1993) 53,
  [\href{http://xxx.lanl.gov/abs/hep-th/9208059}{{\tt
  hep-th/9208059}}].

\bibitem{DHoker:1995uyv}
  E.~D'Hoker and D.~G.~Gagne,
 Worldline path integrals for fermions with general couplings,
 {\em Nucl.\ Phys.}\ B {\bf 467} (1996) 297,
   [\href{http://xxx.lanl.gov/abs/hep-th/9512080}{{\tt
  hep-th/9512080}}].

\bibitem{Reuter:1996zm}
  M.~Reuter, M.~G.~Schmidt and C.~Schubert,
  Constant external fields in gauge theory and the spin 0, 1/2, 1 path integrals,
 {\em Annals Phys.}\  {\bf 259} (1997) 313,
 [\href{http://xxx.lanl.gov/abs/hep-th/9610191}{{\tt
  hep-th/9610191}}].
  
\bibitem{Bastianelli:2002fv}
  F.~Bastianelli and A.~Zirotti,
  Worldline formalism in a gravitational background,
 {\em Nucl.\ Phys.}\ B {\bf 642} (2002) 372,
  [\href{http://xxx.lanl.gov/abs/hep-th/0205182}{{\tt
  hep-th/0205182}}].

\bibitem{Bastianelli:2002qw}
  F.~Bastianelli, O.~Corradini and A.~Zirotti,
  Dimensional regularization for N=1 supersymmetric sigma models and the worldline formalism,
 {\em Phys.\ Rev.}\ D {\bf 67} (2003) 104009,
 [\href{http://xxx.lanl.gov/abs/hep-th/0211134}{{\tt
  hep-th/0211134}}].

\bibitem{Bastianelli:2005vk}
  F.~Bastianelli, P.~Benincasa and S.~Giombi,
 Worldline approach to vector and antisymmetric tensor fields,
 {\em JHEP} {\bf 0504} (2005) 010,
  [\href{http://xxx.lanl.gov/abs/hep-th/0503155}{{\tt
  hep-th/0503155}}].

\bibitem{Dai:2008bh}
  P.~Dai, Y.~t.~Huang and W.~Siegel,
  Worldgraph Approach to Yang-Mills Amplitudes from N=2 Spinning Particle,
 {\em JHEP} {\bf 0810} (2008) 027,
    [\href{http://xxx.lanl.gov/abs/0807.0391}{{\tt
  arXiv:0807.0391}}].

\bibitem{Bastianelli:2013tsa}
  F.~Bastianelli and R.~Bonezzi,
 One-loop quantum gravity from a worldline viewpoint,
 {\em JHEP} {\bf 1307} (2013) 016,
  [\href{http://xxx.lanl.gov/abs/1304.7135}{{\tt
  arXiv:1304.7135}}].
  
\bibitem{Bastianelli:2013pta}
  F.~Bastianelli, R.~Bonezzi, O.~Corradini and E.~Latini,
  Particles with non abelian charges,
 {\em JHEP} {\bf 1310} (2013) 098,
  [\href{http://xxx.lanl.gov/abs/1309.1608}{{\tt
  arXiv:1309.1608}}].

\bibitem{Bastianelli:2006rx}
  F.~Bastianelli and P.~van Nieuwenhuizen,
  Path integrals and anomalies in curved space,
  Cambridge University Press, Cambridge UK (2006).

\bibitem{Gershun:1979fb}
  V.~D.~Gershun and V.~I.~Tkach,
 Classical And Quantum Dynamics Of Particles With Arbitrary Spin,
 {\em JETP Lett.}\  {\bf 29} (1979) 288,
   [Pisma Zh.\ Eksp.\ Teor.\ Fiz.\  {\bf 29} (1979) 320].

\bibitem{Henneaux:1987cp}
  M.~Henneaux and C.~Teitelboim,
 First and second quantized point particles of any spin,
 Conference: 
 \href{https://inspirehep.net/record/966776}{{\tt
  C87-12-17}}, p.113-152.

\bibitem{Howe:1988ft}
  P.~S.~Howe, S.~Penati, M.~Pernici and P.~K.~Townsend,
  Wave Equations for Arbitrary Spin From Quantization of the Extended Supersymmetric Spinning Particle,
 {\em Phys.\ Lett.}\ B {\bf 215} (1988) 555.

\bibitem{Kuzenko:1995mg}
  S.~M.~Kuzenko and Z.~V.~Yarevskaya,
  Conformal invariance, N extended supersymmetry and massless spinning particles in anti-de Sitter space,
 {\em Mod.\ Phys.\ Lett.}\ A {\bf 11} (1996) 1653,
  [\href{http://xxx.lanl.gov/abs/hep-th/9512115}{{\tt
  hep-th/9512115}}].

\bibitem{Bastianelli:2008nm}
  F.~Bastianelli, O.~Corradini and E.~Latini,
 Spinning particles and higher spin fields on (A)dS backgrounds,
 {\em JHEP} {\bf 0811} (2008) 054,
   [\href{http://xxx.lanl.gov/abs/0810.0188}{{\tt
  arXiv:0810.0188}}].
  
\bibitem{Corradini:2010ia}
  O.~Corradini,
  Half-integer Higher Spin Fields in (A)dS from Spinning Particle Models,
  {\em JHEP} {\bf 1009} (2010) 113
 [\href{http://xxx.lanl.gov/abs/1006.4452}{{\tt arXiv:1006.4452}}].

\bibitem{Bastianelli:2009eh}
  F.~Bastianelli, O.~Corradini and A.~Waldron,
  Detours and Paths: BRST Complexes and Worldline Formalism,
 {\em JHEP} {\bf 0905} (2009) 017,
  [\href{http://xxx.lanl.gov/abs/hep-th/0902.0530}{{\tt
  arXiv:0902.0530}}].

\bibitem{Bastianelli:2012bn}
  F.~Bastianelli, R.~Bonezzi, O.~Corradini and E.~Latini,
  Effective action for higher spin fields on (A)dS backgrounds,
 {\em JHEP} {\bf 1212} (2012) 113,
  [\href{http://xxx.lanl.gov/abs/1210.4649}{{\tt
  arXiv:1210.4649}}].

\bibitem{Segal:2001di}
  A.~Y.~Segal,
  Point particle in general background fields versus gauge theories of traceless symmetric tensors,
  {\em Int.\ J.\ Mod.\ Phys.}\ A {\bf 18} (2003) 4999,
  [\href{http://xxx.lanl.gov/abs/hep-th/0110056}{{\tt
  hep-th/0110056}}].
  
\bibitem{Bonezzi:2012vr}
  R.~Bonezzi, O.~Corradini, S.~A.~Franchino Vinas and P.~A.~G.~Pisani,
  Worldline approach to noncommutative field theory,
  {\em J.\ Phys.}\ A {\bf 45} (2012) 405401,
  [\href{http://xxx.lanl.gov/abs/1204.1013}{{\tt
  arXiv:1204.1013}}].

\bibitem{Ahmadiniaz:2015qaa}
  N.~Ahmadiniaz, O.~Corradini, D.~D'Ascanio, S.~Estrada-Jiménez and P.~Pisani,
  Noncommutative U(1) gauge theory from a worldline perspective,
  {\em JHEP} {\bf 1511} (2015) 069
 [\href{http://xxx.lanl.gov/abs/1507.07033}{{\tt
  arXiv:1507.07033}}].

\bibitem{Craigie:1983fb}
  N.~S.~Craigie, V.~K.~Dobrev and I.~T.~Todorov,
  Conformally Covariant Composite Operators in Quantum Chromodynamics,
 {\em Annals Phys.}\  {\bf 159} (1985) 411.

\bibitem{Berends:1985xx}
  F.~A.~Berends, G.~J.~H.~Burgers and H.~van Dam,
  Explicit Construction of Conserved Currents for Massless Fields of Arbitrary Spin,
 {\em Nucl.\ Phys.}\ B {\bf 271} (1986) 429.

\bibitem{Bekaert:2009ud}
  X.~Bekaert, E.~Joung and J.~Mourad,
  On higher spin interactions with matter,
 {\em JHEP} {\bf 0905} (2009) 126,
 [\href{http://xxx.lanl.gov/abs/0903.3338}{{\tt
  arXiv:0903.3338}}].

\bibitem{Weyl:1927vd}
  H.~Weyl,
  Quantum mechanics and group theory,
 {\em Z.\ Phys.}\  {\bf 46} (1927) 1.

\bibitem{Wigner:1932eb}
  E.~P.~Wigner,
  On the quantum correction for thermodynamic equilibrium,
 {\em Phys.\ Rev.}\  {\bf 40} (1932) 749.
 
\bibitem{Moyal:1949sk}
  J.~E.~Moyal,
  Quantum mechanics as a statistical theory,
 {\em Proc.\ Cambridge Phil.\ Soc.}\  {\bf 45} (1949) 99.
 
\bibitem{Sato:1976hy}
  M.~a.~Sato,
  Operator Ordering and Perturbation Expansion in the Path Integration Formalism,
 {\em Prog.\ Theor.\ Phys.}\  {\bf 58} (1977) 1262.
 
\bibitem{Joung:2015eny}
  E.~Joung, S.~Nakach and A.~A.~Tseytlin,
 Scalar scattering via conformal higher spin exchange,
 {\em JHEP} {\bf 1602} (2016) 125,
 [\href{http://xxx.lanl.gov/abs/1512.08896}{{\tt
  arXiv:1512.08896}}].
  
\bibitem{Beccaria:2016syk}
  M.~Beccaria, S.~Nakach and A.~A.~Tseytlin,
  On triviality of S-matrix in conformal higher spin theory,
 {\em JHEP} {\bf 1609} (2016) 034,
 [\href{http://xxx.lanl.gov/abs/1607.06379}{{\tt
  arXiv:1607.06379}}].
  
\end{thebibliography}

\renewcommand\bibname{References}

\end{document}